\documentclass[a4paper,10pt]{article}
\usepackage{latexsym, amsmath, amsfonts, amssymb, array, dcolumn, hhline, longtable, enumerate,psfrag,bm}
\usepackage[english]{babel}
\usepackage[dvips]{graphicx}
\usepackage{epsfig}
\usepackage[small,bf]{caption}
\begin{document}
%
%       \batchmode
%
% document
%

\noindent {\bf \Large{Chemical functionalization of graphene with defects}}

%\date{\today}
\vspace{0.5 cm}

\noindent {\bf \large {D. W. Boukhvalov and M. I. Katsnelson}}

\vspace{0.5 cm}

\noindent {\it Institute for Molecules and Materials, Radboud
University Nijmegen, Heyendaalseweg 135, 6525 AJ Nijmegen,
~~~~~~~~~ ~~~~~~~~~ ~~~~~The Netherlands}

\begin{abstract}
Defects change essentially not only electronic but also chemical
properties of graphene being centers of its chemical activity.
Their functionalization is a way to modify electronic and crystal
structure of graphene which may be important for graphene-based
nanoelectronics. Using hydrogen as an example, we have simulated a
chemistry of imperfect graphene for a broad class of defects
(Stone-Wales defects, bivacancy, nitrogen substitution impurity,
and zigzag edges) by density functional calculations. We have
studied also an effect of finite width of graphene nanoribbons on
their chemical properties. It is shown that magnetism at graphene
edges is fragile with respect to oxidation and, therefore, a
chemical protection of the graphene edges may be required for
application of graphene in spintronics. At the same time,
hydrogenation of the Stone-Wales defects may be a perspective way
to create magnetic carbon.
\end{abstract}

Graphene \cite{r1,r2,r3} is a novel and very perspective material
for nanoelectronics \cite{r1,science}. Effect of various
imperfections of the crystal lattice, such as Stone-Wales defects
(SW) \cite{SW}, vacancies \cite{vac}, substitution atoms
\cite{N,B}, etc., on physical properties of graphene was studied
theoretically \cite{voz1,voz2,Duplock,Roche}. Recently, first
experimental observations of defects in graphene have been
reported \cite{Meyer,jacsvac,ups}. Earlier, the defects in carbon
nanotubes have been intensively investigated
\cite{def,CNTN1,CNTN2,swnt,Krashrev}. It is known that defects can
influence essentially on chemistry of graphene, and there are
several theoretical considerations of chemical functionalization
of imperfect graphene \cite{Duplock,sssw,cooh}; also, the defects
were experimentally used as attractors for impurities
\cite{jacsvac}.

It is still unclear, however, what are typical defects in real
graphene samples and what is their concentration. Further studies
would be important both scientifically and technologically, to
find a way to produce defect-free graphene or graphene with
desired defects to modify electronic properties, like in
conventional semiconductors. The chemical functionalization is
probably the most suitable way to detect imperfections in graphene
crystal lattice. However, only chemisorption of individual atoms
or their pairs were considered before, whereas it was demonstrated
that, say, for the case of hydrogenation of ideal graphene a
complete coverage turned out to be the energetically most
favorable so, the most probably, impurities would tend to form the
whole clusters on the top of imperfect graphene \cite{H,GO}.
Therefore a systematic study of chemisorption on defects in
graphene for a broad range of coverage is desirable; this is the
subject of the present work.

Edges of graphene sheets which are broadly discussed now, due to
their potential relevance for field-effect transistors and
spintronic devices \cite{tranz1,tranz2,Louie,Kaxiras,YK,Yang, kit2} can
be also considered as defects. Their functionalization and its
effect on the electronic structure were discussed in a series of
theoretical works
\cite{Yang,Ferarri2,Hod1,Hod2,Hod3,Hod4,Huang,magnonano,kit1,O2},
chemical activity of the graphene edges being discussed
\cite{reakt}, but only the chemical functionalization of the edges
themselves were considered rather than of the whole graphene sheet
{\it with} the edges. Actually, the edges change chemical activity
of graphene in an extended region, as will be demonstrated here.
This is important to study stability of magnetism at the edges and
to evaluate a width of a ``near-edge layer'' where properties of
graphene are essentially different from those in the bulk.

We will use hydrogenation as an example of the chemical
functionalization of defects in graphene. For the case of ideal
graphene general factors determining the chemisorption were first
studied for the case of hydrogen \cite{H} and appeared to be
applicable also in other situations, such as graphene oxide
\cite{GO}. One can expect therefore that, at least, some of
peculiarities which are found here can be of a more general
relevance.

We used the pseudopotential density functional SIESTA package for
electronic structure calculations \cite{siesta1, siesta2} with the
generalized gradient approximation for the density functional
\cite{PBE}, with energy mesh cutoff 400 Ry, and $k$-point
8$\times$8$\times$2 mesh in Monkhorst-Park scheme \cite{MP} for
modeling of defects, and 2$\times$12$\times$2 for modeling of
nanoribbons. During the optimization, the electronic ground state
was found self-consistently by using norm-conserving
pseudopotentials for cores and a double-$\zeta$ plus polarization
basis of localized orbitals for carbon and oxygen, and
double-$\zeta$ one for hydrogen. Optimization of the bond lengths
and total energies was performed with an accuracy 0.04 eV /\AA
~and 1 meV, respectively. This method is frequently used for
computations of electronic structure of graphene
\cite{swform,Louie,Kaxiras,H,GO}. When drawing the pictures of
density of states, a smearing by 0.2 eV was used.

\begin{figure}[ht]
 \begin{center}
   \centering
\includegraphics[width=5.2 in]{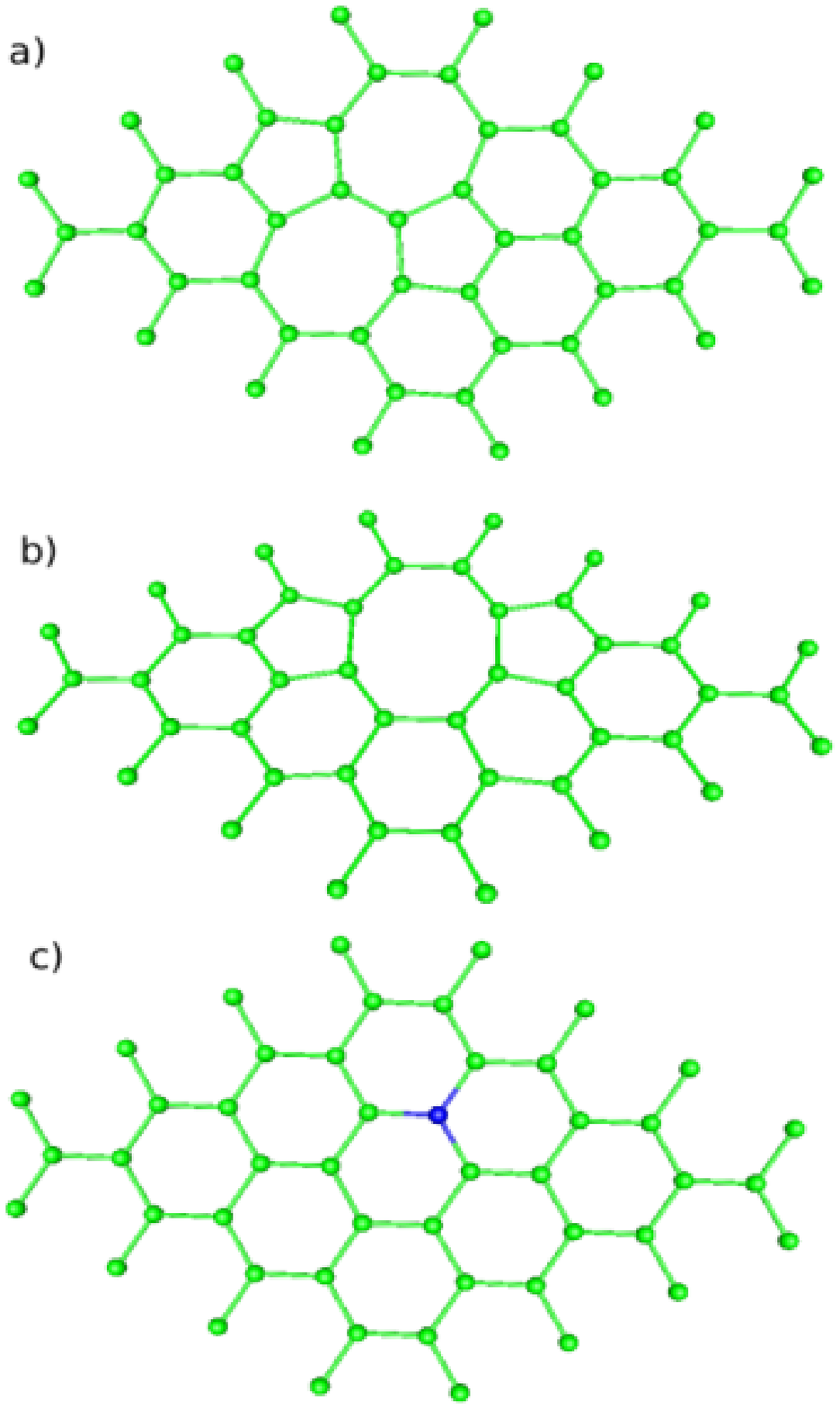}
\caption{Optimized geometric structures for graphene supercell
containing the Stone-Wales defect (a), bivacancy (b), and nitrogen
substitution impurity (c).}
            \label{fig1}
 \end{center}
\end{figure}

\begin{figure}[ht]
 \begin{center}
   \centering
\includegraphics[width=5.2 in]{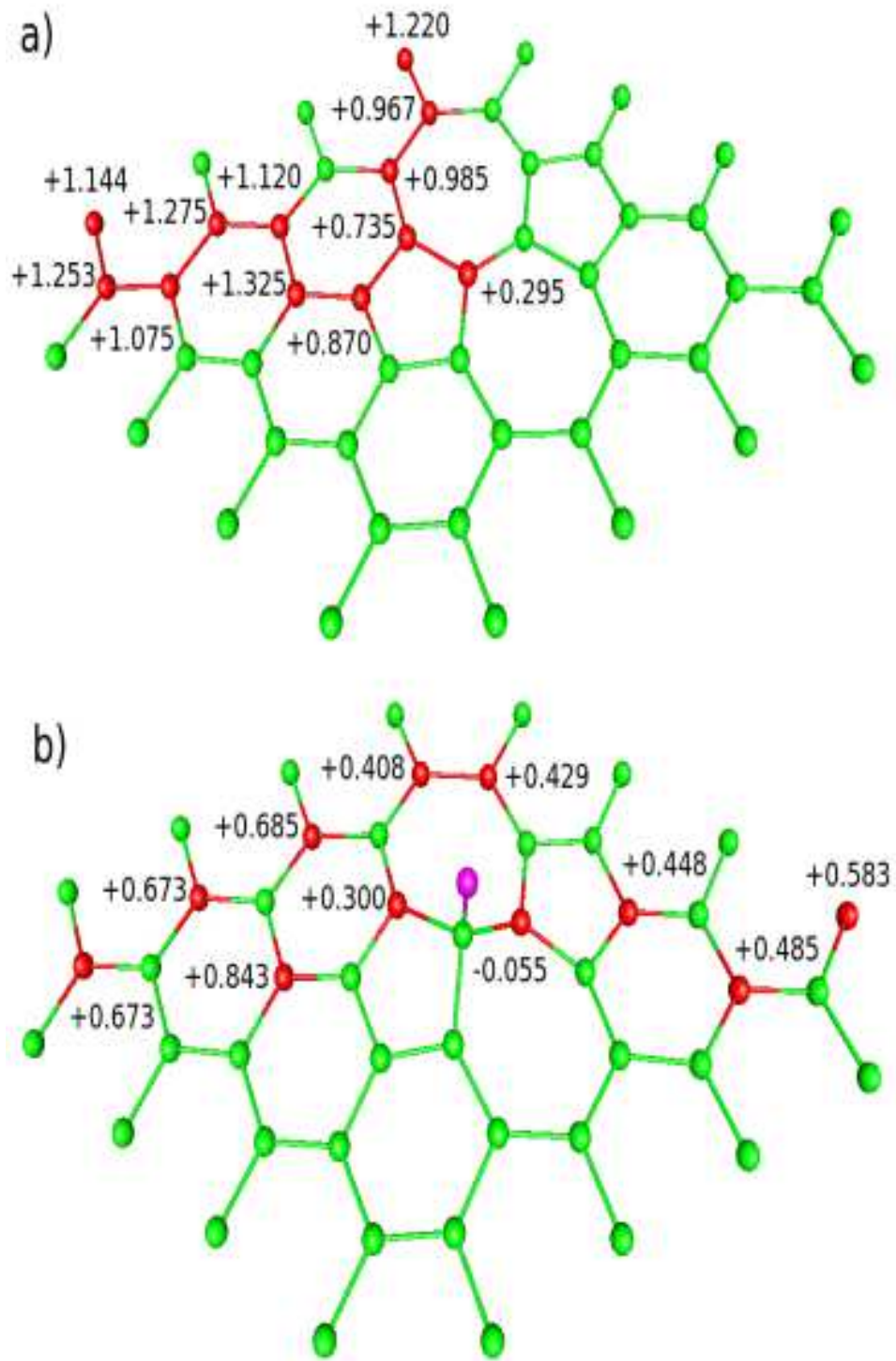}
\caption{Formation energies (in eV per hydrogen atom) for
nonequivalent sites (red circles) of graphene lattice with the SW
defect for chemisorption of the first (a) and second (b) hydrogen
atoms.}
            \label{fig2}
 \end{center}
\end{figure}

Chemisorption energies were calculated by standard formulas used,
e.g., earlier for the cases of chemisorption of hydrogen \cite{H}
oxygen and hydroxyl groups \cite{GO} on graphene and solution of
carbon in $\gamma$-iron \cite{iron}. Thus, energy of the
chemisorption of single hydrogen atom at graphene with Stone-Wales
defect (Fig. \ref{fig1}a) was calculated as
E$_{form}$=E$_{SW+H}$-E$_{SW}$-E$_{H_2}$/2 where E$_{SW+H}$ was
the total energy of the supercell with chemisorbed hydrogen found
by self-consistent calculations after optimization of geometric
structure, E$_{SW}$ was the total energy of the graphene supercell
with the Stone-Wales defect found in a similar way, and E$_{H_2}$
is the energy of hydrogen molecule. The latter choice describes
the most adequately a process of spontaneous hydrogen dissociation
and, thus, stability of graphene with respect to the
hydrogenation. That is why this definition is broadly used when
discussing hydrogen storage \cite{Dillon1, Dillon2}. In principle,
a definition of the hydrogenation energy depends on what specific
compound is used for the hydrogenation. This may shift the
chemisorption energy values of hydrogen in Figs. \ref{fig2},
\ref{fig4}, \ref{fig9}, and \ref{fig10}. For example, for the
hydrogenation by hydrogen plasma all formation energies presented
below will be lower by 104.206 kcal/mol, or 2.255 eV per atom,
which is the experimental bond dissociation energy for H$_2$.

We used graphene supercell containing 50 carbon atoms, similar to
that used in our previous work \cite{H}. It was demonstrated there
that the chemisorption energy for two hydrogen atoms is close to
that for two independent single atoms if the distance between the
atoms is larger than 1 nm (approximately, four graphene lattice
constants). Similar results have been obtained for the case of
hydroxy groups \cite{GO}. This means that a supercell provided the
distance between defects larger than 1 nm is sufficient to model
the chemisorption on individual defects.

Optimized geometric structures of graphene supercells with various
kinds of defects are presented in Fig. \ref{fig1}. To simulate
graphene nanoribbons we used the stripes with 22, 44, and 66
carbon atoms per cell, with the width 2.20, 4.54, and 6.88 nm,
respectively.

We start with the calculations of activation energy, that is, the
chemisorption of a single hydrogen atom. For pure graphene this
energy is 1.5 eV \cite{H} and for all kind of defects under
consideration it turns out to be lower, namely, 0.30 eV for SW,
0.93 eV for bivacancy, and 0.36 eV for substitution impurity of
nitrogen (in the latter case we mean the energy of chemisorption
on a carbon neighbor of the nitrogen impurity).

We have calculated also the chemisorption energy for different
nonequivalent atoms of the lattice containing the SW defect (Fig.
\ref{fig2}a). One can see that the chemisorption energy at the SW
defect is minimal. At the same time, in a whole area surrounding
the defect the chemisorption energy is lower than in perfect
graphene \cite{H}. For distant enough atoms the energy is very
close to the values found in Ref. \cite{H} which confirms that the
supercell with 50 atoms is sufficient to describe the
chemisorption on the single defect.

\begin{figure}[ht]
 \begin{center}
   \centering
\includegraphics[width=5.2 in]{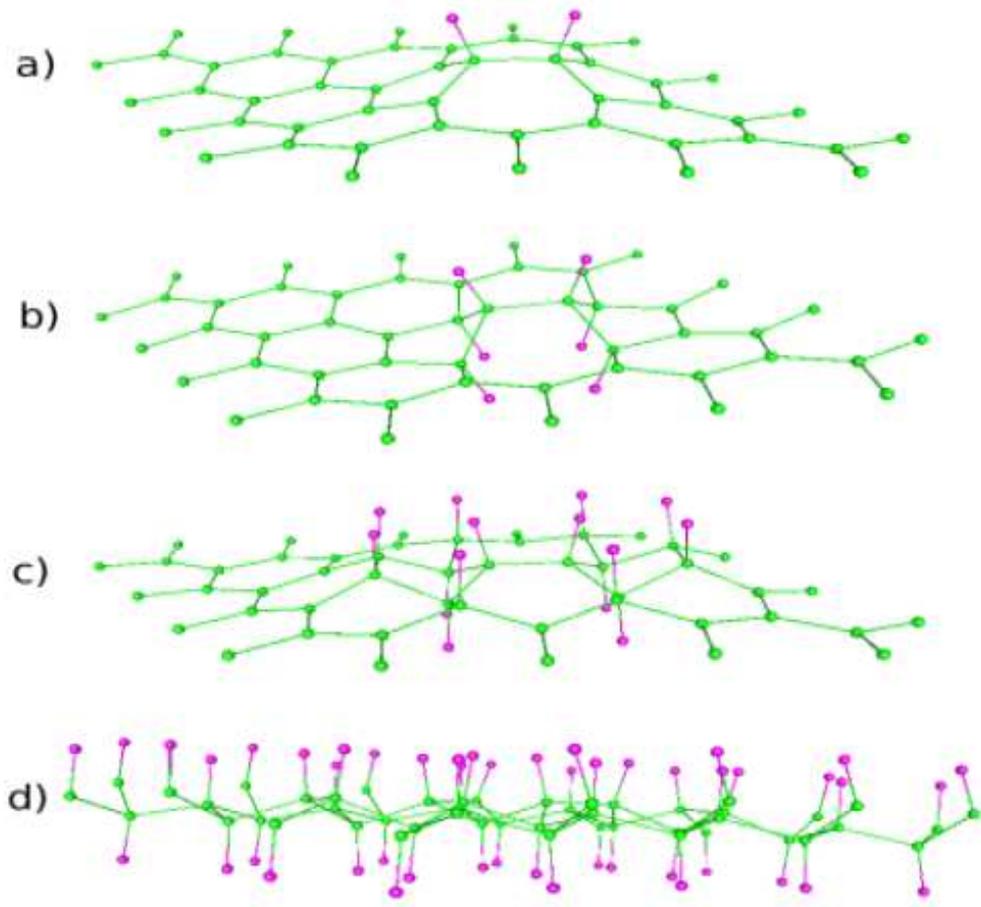}
\caption{Optimized geometric structures for the Stone-Wales defect
functionalized by 2 (a), 6 (b), 14 (c) hydrogen atoms and
completely covered by hydrogen (d). Carbon and hydrogen atoms are
shown by green and violet circles, respectively.}
            \label{fig3}
 \end{center}
\end{figure}

\begin{figure}[ht]
 \begin{center}
   \centering
\includegraphics[width=5.2 in]{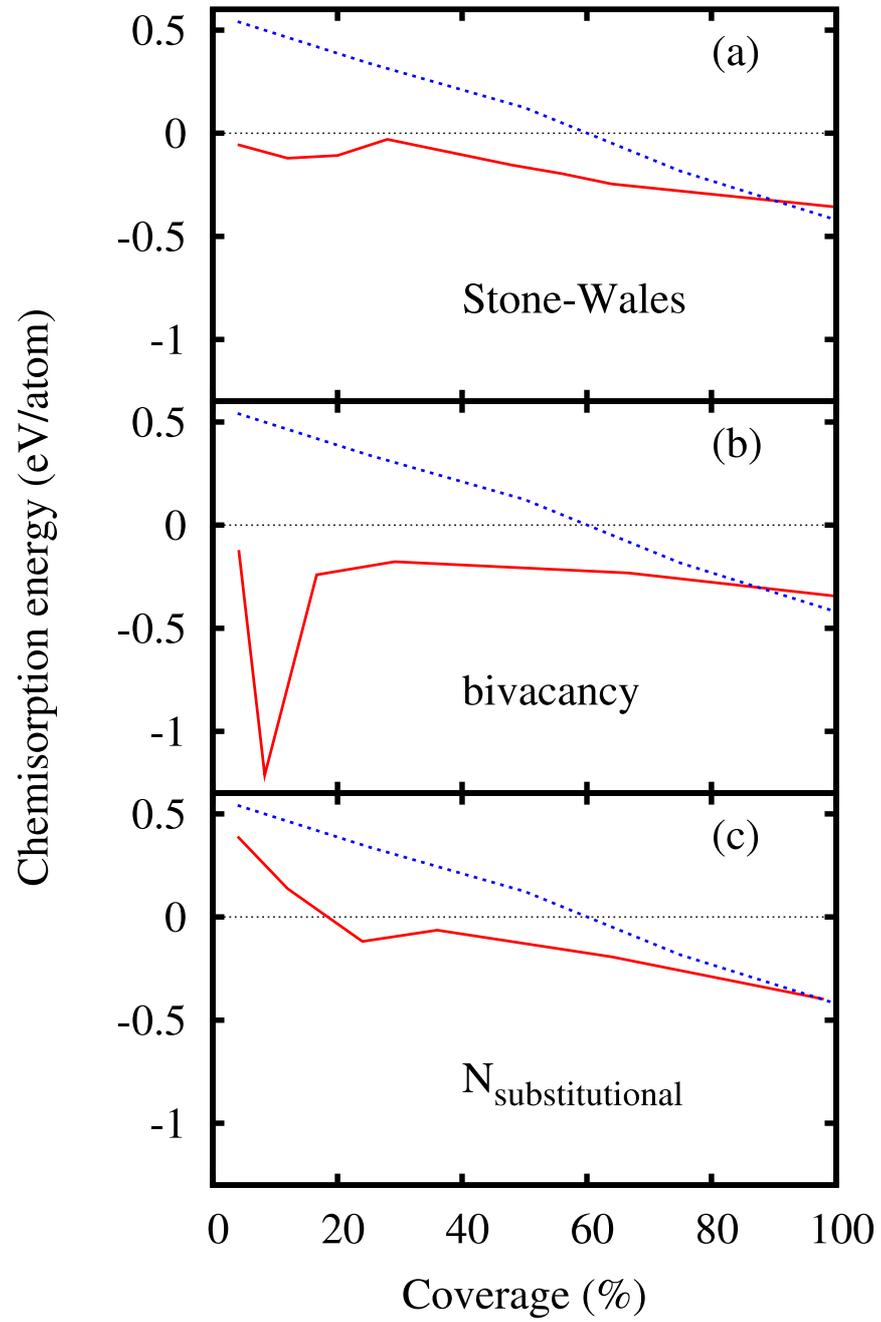}
\caption{Energy formation as a function of coverage for grapehne
sheet containing the Stone-Wales defect (a), bivacancy (b), and
nitrogen substitution impurity (c). The blue dashed line
represents the results for the ideal infinite graphene sheet.}
            \label{fig4}
 \end{center}
\end{figure}

The computational results mean that defects in graphene are
centers of chemical activity. As a next step, we investigate an
interaction of a pair of hydrogen atoms with defects (Fig.
\ref{fig3}a). Again, the defects decrease the chemisorption energy
of the pair in comparison with the case of ideal graphene. We have
calculated also the pair formation energy for the case when the
first hydrogen atom is chemisorbed on the SW defect and the second
one - on one of the surrounding carbon atoms (Fig. \ref{fig2}b).
The chemisorption energies are presented there for the case of
one-side chemisorption; for the first and second neighbors of the
SW defect the one-side chemisorption is energetically more
favorable \cite{Duplock,sssw,cooh} whereas for farther atoms the
difference with the two-side chemisorption energies becomes
negligible \cite{H}. Similar to the case of single hydrogen atom
discussed above a region is formed around the defect with a
diameter of order of 1 nm where the chemisorption energies are
lower than in the bulk.

These results may be essential for the problem of magnetism in
graphene. Chemisorption of a single hydrogen atom on graphene
results in appearance of magnetic moments \cite{Duplock,Krash,XPS}
whereas close hydrogen pairs on graphene turn out to be
nonmagnetic, the pairs being more energetically favorable than
independent atoms \cite{H,jap1,jap2}. At the same time, there is
an energy barrier which two distant hydrogen atoms on graphene
should overcome to form the pair \cite{barier}. If one of these
atoms are bonded with the SW defect the barrier becomes higher by
0.44 eV which makes the pair formation at room temperature rather
difficult. Thus, the hydrogenation of the SW defects may be a
perspective way to create magnetic graphene. It can be done, for
example, using reactions with superacids, similar to the case of
fullerenes \cite{superacid}.

For SW and bivacancies the chemisorption energy of single hydrogen
atom is negative which makes these defects centers of {\it stable}
chemisorption of hydrogen. At further increase of the number of
hydrogen atoms the chemisorption energy decreases reaching a
minimum for the most stable configuration (for the case of SW it
is shown in Fig. \ref{fig3}b), then increases up to a local
maximum (the corresponding structure is shown in Fig. \ref{fig3}c)
and then decreases again, with the global minimum for complete
coverage. Contrary, in the case of ideal graphene the
chemisorption energy decreases monotonously with the hydrogen
concentration and clusters of chemisorbed hydrogen are not stable
\cite{clust}. Interestingly, for the case of complete coverage the
binding energy is smaller for the hydrogen on graphene with
defects than on the perfect one. This means that completely
hydrogenated graphene (graphane \cite{H,sofo}) is less stable with
defects than without them.

Our results show that, whereas at the hydrogenation of pure
graphene there is single potential barrier to overcome, that is,
the adsorption of the first atom, for imperfect graphene several
barriers can exist (see Fig. \ref{fig4}). The first one, again,
corresponds to the activation energy and the other ones - to the
transition from hydrogenation of the region around defect with the
minimal chemisorption energy (see Fig. \ref{fig2}) to the
hydrogenation of the whole graphene sheet. Studying experimentally
kinetics of this process one can extract an information about
presence of defects in graphene and their type. At the same time,
graphene with defects can be a good model of activated carbon
\cite{active1,active2}. The fact that the adsorption and
desorption energies are smaller than 0.5 eV per hydrogen atom
makes graphene with defects a perspective catalyst.

\begin{figure}[ht]
 \begin{center}
   \centering
\includegraphics[width=5.2 in]{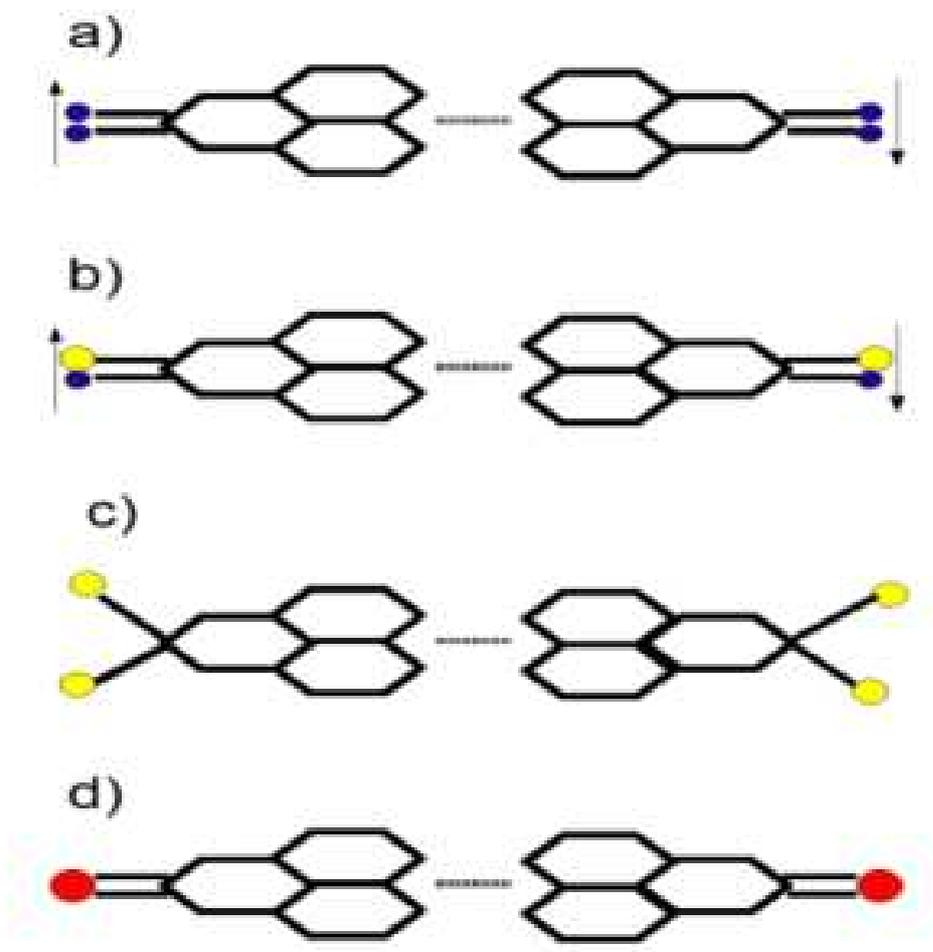}
\caption{A sketch of functionalization of the zigzag edges of
graphene: (a) Initial state of edges, unpaired electrons are shown
in blue; (b), (c), and (d) the edge functionalized by a single
carbon atom (yellow), by pair of hydrogen atoms, and by oxygen
atom (red), respectively. }
            \label{fig5}
 \end{center}
\end{figure}

\begin{figure}[ht]
 \begin{center}
   \centering
\rotatebox{-90}{
\includegraphics[height=5.2 in]{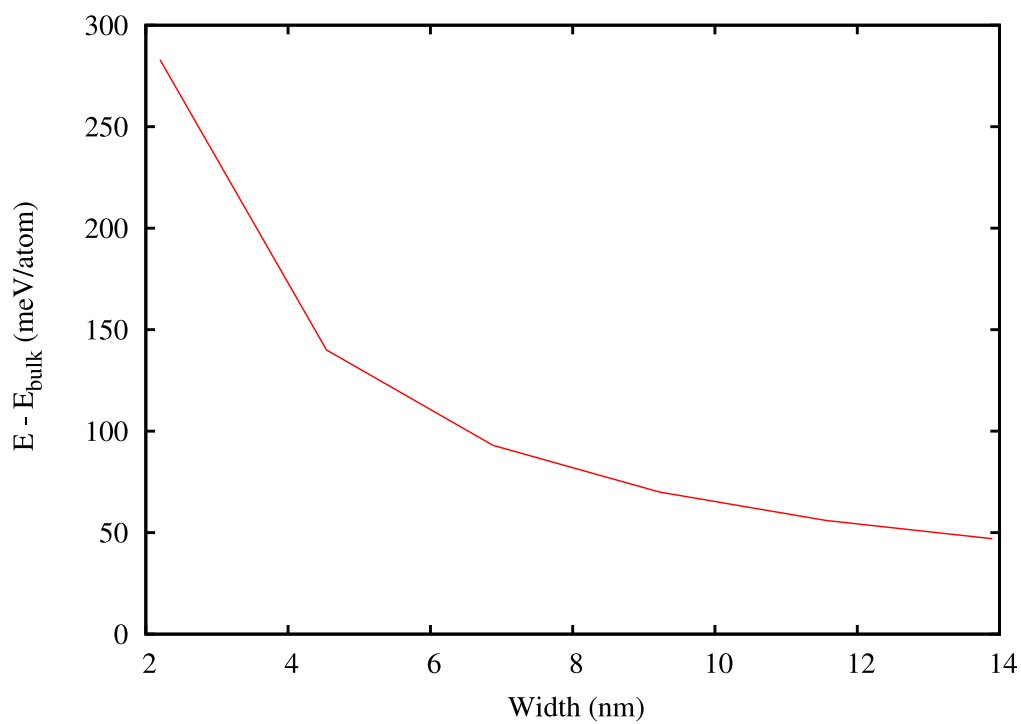}}
\caption{Total energy difference between infinite graphene sheet
and graphene nanoribbon as a function of the nanoribbon width.}
            \label{fig6}
 \end{center}
\end{figure}

\begin{figure}[ht]
 \begin{center}
   \centering
\includegraphics[width=5.2 in]{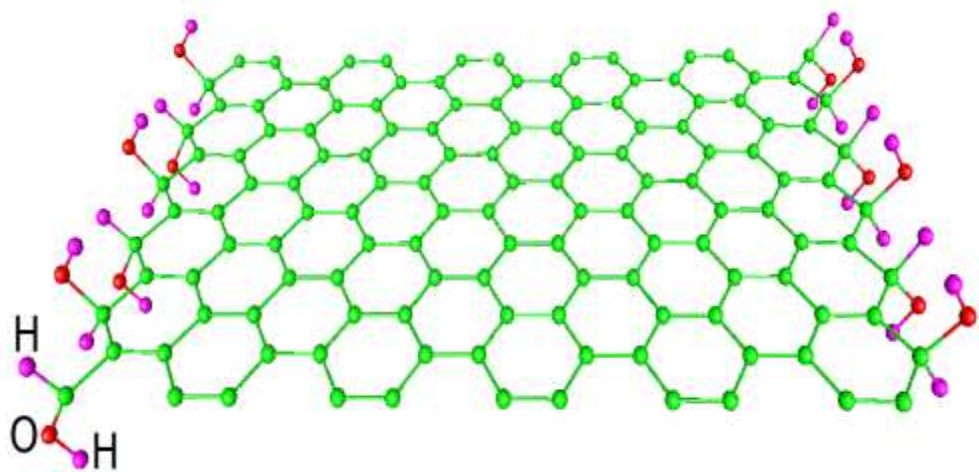}
\caption{Optimized geometry of the graphene nanoribbon of width
2.20 nm with the edges functionalized by hydrogen atoms and
hydroxy groups resulted from water decomposition.}
            \label{fig7}
 \end{center}
\end{figure}

Let us consider now the hydrogenation of graphene edges. We focus
on the zigzag edges which are especially interesting due to their
possible half-metallic ferromagnetic state
\cite{Louie,Kaxiras,YK}. For the nanoribbons of width 2.20, 4.54,
and 6.88 nm, the chemisorption energies for a single hydrogen atom
(Fig. \ref{fig5}b) are found to be -2.39, -2.76, and -2.82 eV,
respectively. High chemical activity of the edges, in comparison
with the bulk graphene \cite{H} is explained by a presence of
unpaired electrons due to broken bonds (Fig. \ref{fig5}a). Only
one unpaired electron is involved in the bond with the single
hydrogen atom whereas the second one can participates in the
second bond (Fig. \ref{fig5}c). The chemisorption of the second
hydrogen atom leads to strong local distortions of graphene, due
to a change of hybridization from sp$^2$ (favorable for a planar
geometry) to sp$^3$ (favorable for three-dimensional structure
like in diamond). These distortions require some energy which is
higher than the energy gain due to pairing of all electrons and,
as a result, an energy barrier about 1.5 eV is formed for all
nanoribbons under investigation. The cohesive energy of pure
graphene is rather strongly dependent on the ribbon width (Fig.
\ref{fig6}). This result in a strong dependence of the
chemisorption energy on the ribbon width as well. The energy of
chemisorption of two hydrogen atoms at the edges is negative, in
contrast with the case of perfect bulk graphene \cite{H} which
makes the edges another natural centers of chemical activity.
Similarly, zigzag edges can adsorb two fluorines \cite{F}, or two
hydroxyl groups, or a pair of hydroxyl group and hydrogen, the
latter case being related to the decomposition of a water molecule
(Fig. \ref{fig7}). The formation energy per H$_2$O molecule
corresponding to this process is -2.42 eV (241 kJ/mol) for the
ribbon width 2.20 nm which makes the decomposition of the water
more favorable than the chemisorption of single hydrogen atom.

In a case of chemisorption of oxygen atoms, both unpaired
electrons are involved but the hybridization remains sp$^2$ and
local geometry therefore remains flat (Fig. \ref{fig5}d). The
chemisorption energy for the oxygen atom is much lower than for
hydrogen or hydroxyl; for the ribbon width 2.2 nm it turns out to
be -4 eV. Recently, peaks corresponding to C=O chemical bonds have
been found in the core-level spectra of carbon in graphene
nanoribbons \cite{ribgrow}. They were related to edge groups,
based on ratio of intensities of these peaks and of the main peaks
corresponding to sp$^2$ hybridization.

\begin{figure}[ht]
 \begin{center}
   \centering
\includegraphics[width=5.2 in]{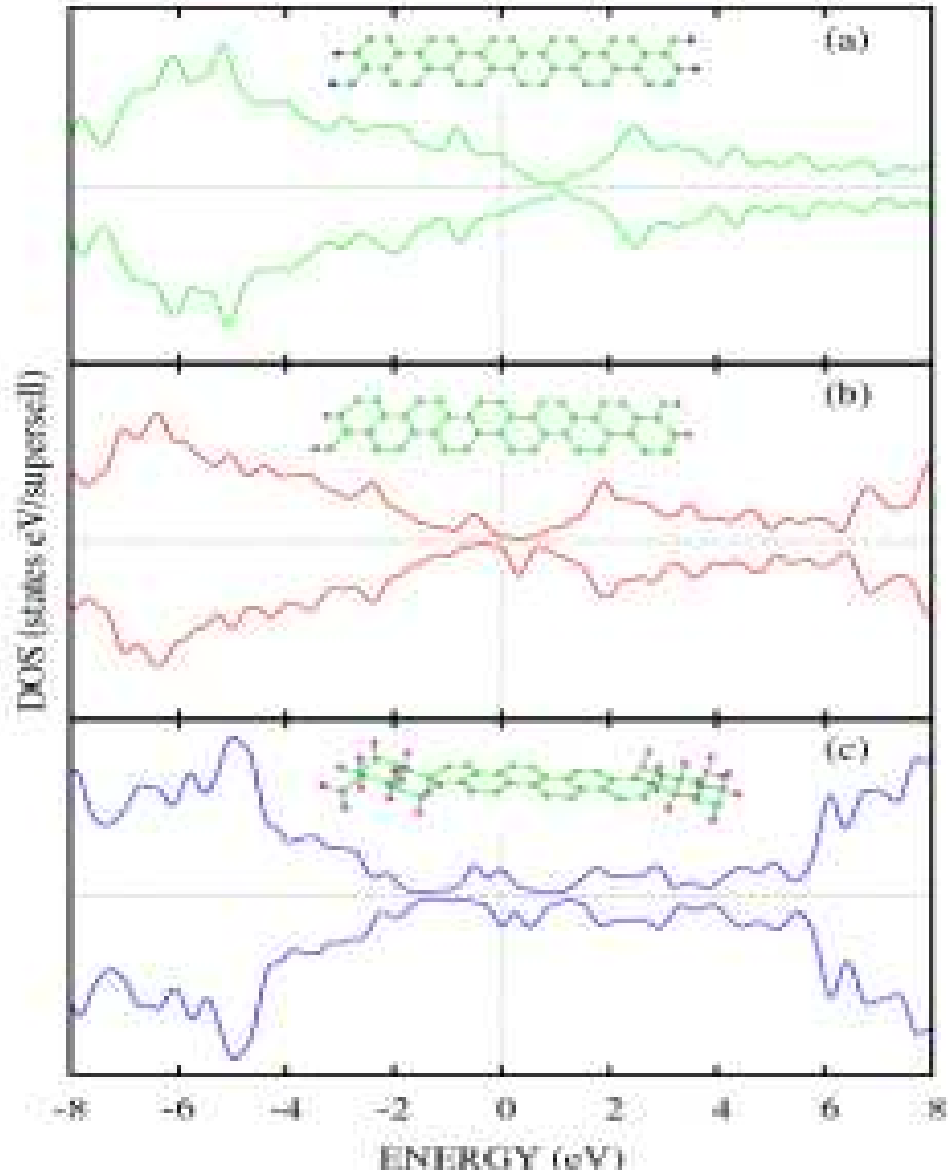}
\caption{Densities of states for graphene nanoribbon of width of
2.2 nm with its edges functionalized by: (a) oxygen atoms and (b)
hydrogen atoms;
(c) the case of complete coverage of six first layers from the edges
 On insets optimized geometric structures of fuctionalized nanoribbons
corresponding with dencities of states.}
            \label{fig8}
 \end{center}
\end{figure}

\begin{figure}[ht]
 \begin{center}
   \centering
\includegraphics[width=5.2 in]{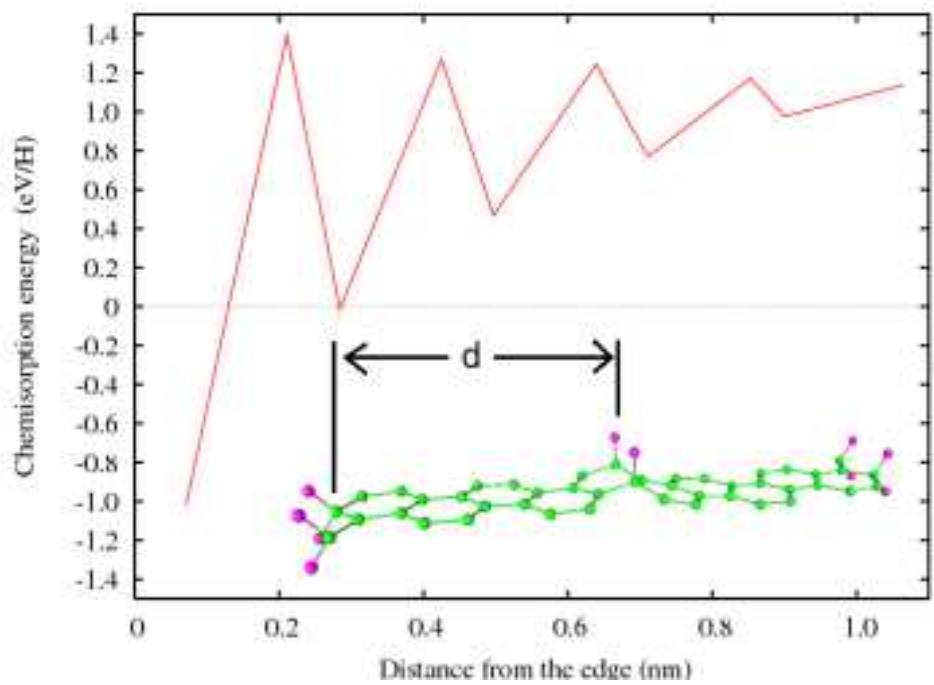}
\caption{Chemisorption energy of hydrogen as a function of
distance $d$ from the edge. Inset: optimized geometric structure
of the system under consideration.}
            \label{fig9}
 \end{center}
\end{figure}

\begin{figure}[ht]
 \begin{center}
   \centering
\includegraphics[width=5.2 in]{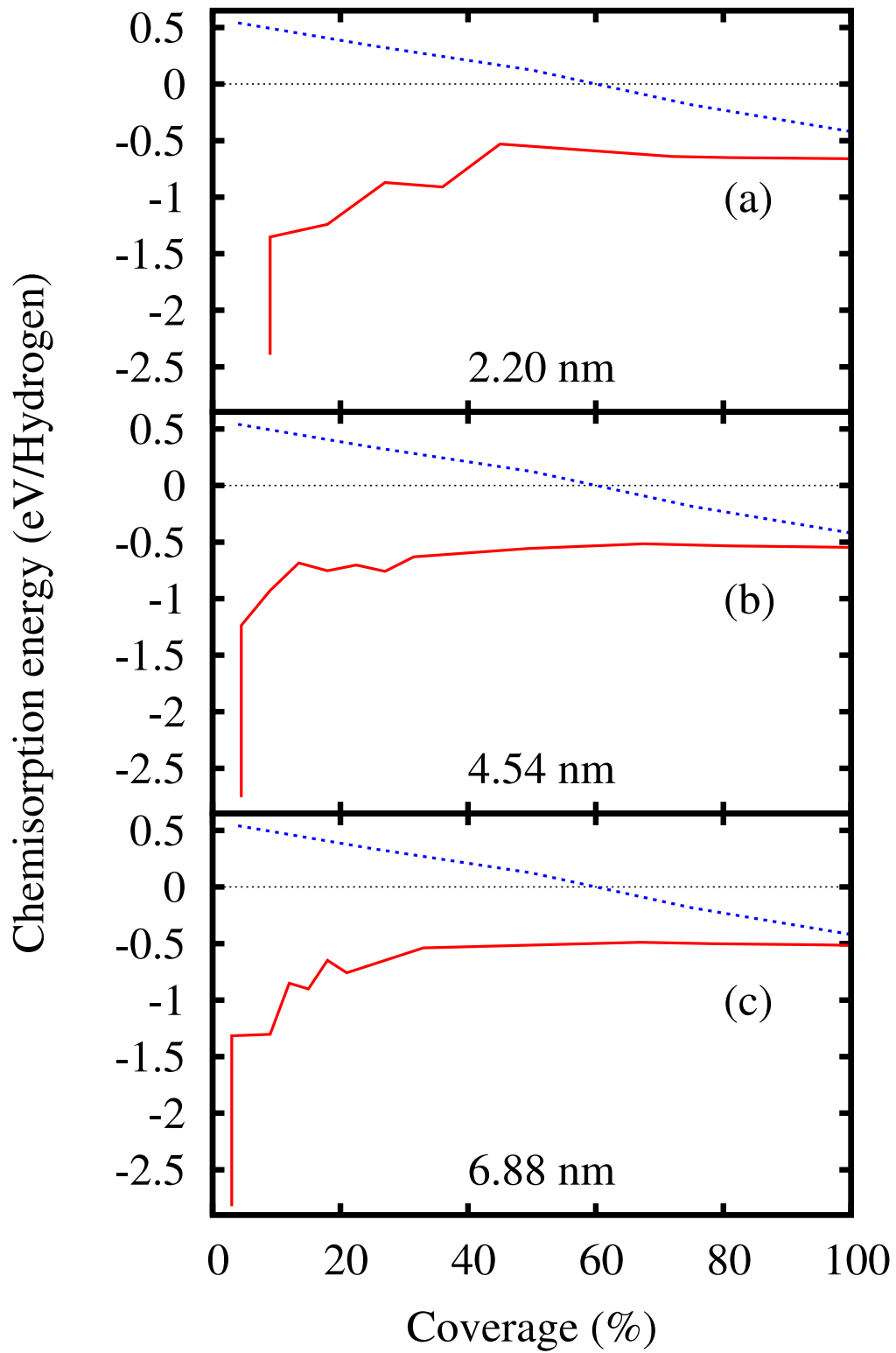}
\caption{Formation energy as a function of coverage of graphene
nanoribbons by hydrogen as a function of the nanoribbon width (red
solid line). The blue dashed line presents the results for an
ideal infinite graphene sheet.}
            \label{fig10}
 \end{center}
\end{figure}

Since all unpaired electrons are bonded at the complete
oxygenation of the edge, the magnetism disappears and electronic
structure becomes purely metallic (see Fig. \ref{fig8}), instead
of half-metallic ferromagnetic for undoped zigzag edges
\cite{Louie}. Note that the decomposition of the water molecule
described above will also suppress the magnetism of the zigzag
edges, as was observed at fluorination of the graphene edges
\cite{F}.

Now we consider further hydrogenation of graphene, from the edges
to bulk. The computational results presented in Fig. \ref{fig9}
show that the chemisorption of hydrogen is the most energetically
favorable near the edges and less favorable in the middle of the
nanoribbon where the formation energy is close to that for the
bulk (1.44 eV \cite{H}). The chemisorption energy is very
sensitive to the distortions created by adsorbed hydrogen atoms.
It was shown earlier for the bulk graphene \cite{H} that the
difference in chemisorption energies for different configurations
of a pair of hydrogen atoms can be as large as 0.45 eV. In
graphene nanoribbons these effects seem to be even stronger. In
particular, geometric frustrations due to the atomic distortions
lead to a zigzag dependence of the chemisorption energy on atomic
position (Fig. \ref{fig9}). At the chemisorption of single
hydrogen atom interactions between the hydrogen-induced
distortions and those due to edge itself is important. This effect
decays when moving from the edge at distances of order of 1 nm. As
well as in the case of SW defect, the pair formation can be more
difficult for the nanoribbon than for bulk graphene, due to larger
barrier energy. This may be essential, again, for the problem of
hydrogen-induced magnetism in graphene.

The results are shown in Fig. \ref{fig10}. One can see that,
irrespective to the ribbon width, at first the formation energy
grows by steps which corresponds to a preferable hydrogenation of
graphene by pairs of atoms \cite{H}. This growth continues at the
depth of 8 carbon atoms (0.85 nm, which is slightly larger than
for the single atom due to two-side hydrogenation) and then the
energy increases approximately linearly with the coverage, as in
the case of infinite graphene sheet \cite{H}. This means that the
hydrogenation of graphene proceeds in three steps, namely, the
hydrogenation of the edge, then, of the near-edge layer with a
width of order 1 nm, and, at last, of the bulk. The chemical
functionalization of the near-edge layer effects essentially on
the electronic structure of graphene nanoribbons (see Fig.
\ref{fig8}). It is worth to note that for the ribbons under
consideration the hydrogenation energy is negative which makes
them chemically active in their whole area. This can influence the
electronic structure as was mentioned earlier \cite{tranz1}.

To conclude, let us summarize the main results. Defects in
graphene crystal lattice, as well as the edges of graphene, change
drastically a scenario of its chemical functionalization. For the
ideal graphene, only single energy barrier related with the
chemisorption of the first atom is relevant and complete
functionalization corresponds to the global energy minimum. At the
same time, for ``realistic'' graphene with imperfections and edges
local energy minima are formed corresponding to local
functionalization of a region with radius of order half of
nanometer and further functionalization is stopped by presence of
another energy barrier. The local energy minima make migration of
chemisorbed atoms and groups more difficult. In the cases of
hydrogen and other atoms and groups with single unpaired electron,
this can stabilize a magnetic state preventing formation of
nonmagnetic pairs. We have found that magnetic state of edges in
graphene nanoribbons may be unstable with respect to oxidation and
water dissociation at the edges. Therefore, keeping in mind
potential applications of this magnetism in spintronics special
methods of chemical protection of the edges should be developed.

{\bf Acknowledgment} ~~ The work is financially supported by
Stichting voor Fundamenteel Onderzoek der Materie (FOM), the
Netherlands.

\end{document}